# Challenges towards Building an effective Cyber Security Operations Centre

**Cyril Onwubiko[*] and Karim Ouazzane[+]**

[*]*Artificial Intelligence, Blockchain and Cyber Security, E-Security Group, Research Series, London, UK*
[+]*Cyber Security Research Centre (CSRC), London Metropolitan University, London, UK*

**ABSTRACT**
The increasing dependency of modern society on IT systems and infrastructures for essential services (e.g. internet banking, vehicular network, health-IT, etc.) coupled with the growing number of cyber incidents and security vulnerabilities have made Cyber Security Operations Centre (CSOC) undoubtedly vital. As such security operations monitoring is now an integral part of most business operations. SOCs (used interchangeably as CSOCs) are responsible for continuously and protectively monitoring business services, IT systems and infrastructures to identify vulnerabilities, detect cyber-attacks, security breaches, policy violations, and to respond to cyber incidents swiftly. They must also ensure that security events and alerts are triaged and analysed, while coordinating and managing cyber incidents to resolution. Because SOCs are vital, it is also necessary that SOCs are effective. But unfortunately, the effectiveness of SOCs are a widespread concern and a focus of boundless debate. In this paper, we identify and discuss some of the pertinent challenges to building an effective SOC. We investigate some of the factors contributing to the inefficiencies in SOCs and explain some of the challenges they face. Further, we provide and prioritise recommendations to addressing the identified issues.

# 1   INTRODUCTION

A Cyber Security Operations Centre (CSOC) is an *essential business function* and should arguably be an integral part of all modern business operations and national cyber security programmes regardless of scale and size. SOCs are responsible for cyber security incident management, cyber-attack detection, continuous and protective security monitoring, log and event management, coordination and investigation (Onwubiko, C., & Ouazzane, K., 2019a).

An effective SOC comprises three key aspects:
- *Building of the central log collection, aggregation, analysis and incident management platform* (a.k.a. **SOC Monitoring Platform**).
- *Onboarding of both new and existing services for continuous monitoring by the SOC monitoring platform* (a.k.a. **Cyber Onboarding**).
- *Performing the continuous monitoring by SOC Analysts through technology, tools and processes (a.k.a. 'Eyes-on-Glass').*

While a *SOC monitoring platform* may be built but the problem lays with onboarding services into it so that they can be continuously and protectively monitored. We use the analogy of a *building* and its *content*. You could have an unfurnished property, where the property is built with the necessary doors and windows, but the property is empty and has no content, such as beds, chairs, cooker or electricity. The same can be said of a *SOC monitoring platform* without onboarding of the services and infrastructures it was built to monitor. Therefore, to have a functioning and operational SOC, then onboarding of services, systems and network infrastructure to the *SOC monitoring platform* must occur (Onwubiko, C. and Ouazzane, K., 2019a).

Further, and equally as important, is meeting business requirements, especially for a multitenant SOC or a SOC monitoring business services of varied security and operational requirements. SOC is not a one-size fits all. It must be tailored, however slightly, to meet unique business and operational use cases.

Unfortunately, many SOCs are believed to be ineffective. According to Schinagl, S., et al., (2015), only a few SOCs are effective in countering cybercrime and IT abuse.

Our contributions in this paper are:
1. This is an extension of our paper, titled "Cyber Onboarding is Broken" (Onwubiko, C. and Ouazzane, K., 2019a), which focusses primarily on 'Cyber Onboarding' – a core component of the SOC. In that work, we utilised the reframing matrix methodology to identify and discuss four stakeholder views in understanding and reasoning how best a SOC can be established to achieve its overarching purpose. Conversely, in this paper, significant extensions have been made, including SOC monitoring and incident management, process, people and technology.
2. We discuss several SOC operating models and their features.
3. We investigate some of the factors contributing to the inefficiencies in SOCs, and finally,
4. We explain some of the challenges SOCs face and how best to address these issues and concerns.

The remainder of the paper is organised as follows: Section 2 provides a detailed review of SOCs, covering SOC vs. SIEM, Insource vs. Outsource, Cloud SOC, On-Premise and Hybrid SOCs. The business requirements of SOCs are presented in Section 3. In Section 4, Cyber Onboarding is briefly discussed, while Section 5 outlines factors contributing to SOC inefficiencies. Challenges facing organisational and national SOCs are explained in Section 6. Section 7 offers recommendations and conclusions, and suggestions for future work in Section 8.

## 2    SECURITY OPERATIONS CENTRE

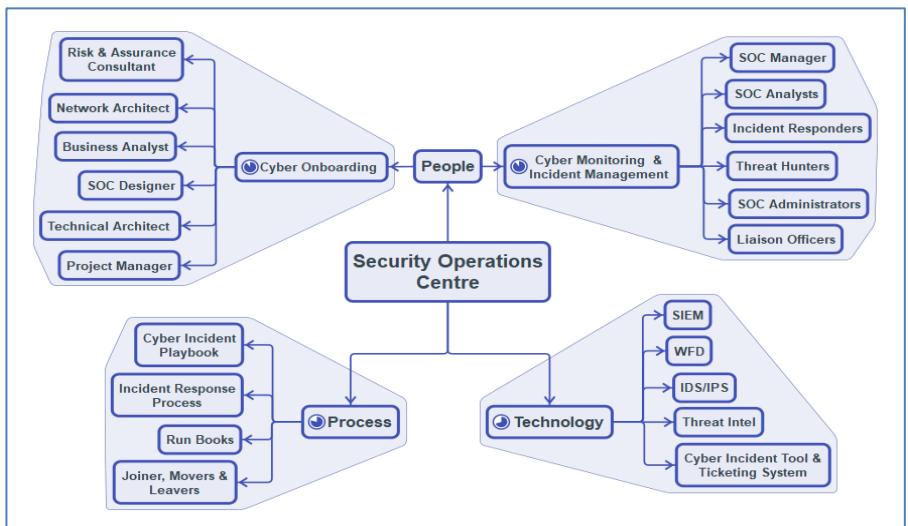

FIG. 1: SOC FUNCTIONAL CONCEPTUAL DIAGRAM

Many people conflate SOC with SIEM. A SIEM (Security Information and Event Management) is a tool, which offers log management, event and log correlation, analysis and dashboard. Conversely, SOC is a business operations function comprising *People, Process* and *Technology* as shown in Fig. 1.

Fig. 1 is a conceptual representation of a SOC, showing the building blocks - *people*, *process* and *technology*. It is pertinent to note that the list of processes, or types of technologies, or categories of people shown in Fig. 1 is deliberately inexhaustive.

## 2.1 People

*People* comprises analysts, administrators, incident responders, SOC manager etc. who perform continuous monitoring (a.k.a. 'eyes-on-glass') of the organisation's business services and IT estate by leveraging the capabilities offered by *Technology* e.g. SIEM tool, and guided by the organisation's *policies, processes and procedures*. So, a SIEM is not a SOC. Rather, a SIEM is only a technology constituent part of a SOC. People can be subdivided into two broad categories: **cyber onboarding** people, and **SOC monitoring and incident management** personnel.

### 2.1.1 Cyber Onboarding

Cyber onboarding is a multidisciplinary team composed of solutions and technical architects, SOC designers, SOC content engineers, business analysts, risks and information assurance consultant and project managers (see Fig. 1). These are the people who carry out *project related activities* to ensuring that each business service (a *business service* usually comprises, at the least, systems, network infrastructures and applications) to be monitored is properly onboarded to a SOC monitoring and incident management platform.

### 2.1.2 SOC Monitoring and Incident Management

SOC monitoring and incident management are solely responsible for security monitoring, continuous and protective monitoring of onboarded services that are in the SOC platform, providing 'eyes-on-glass' monitoring[1], vulnerability scanning, alerting and event analysis, incident triage, cyber incident management, coordination and reporting. They are also the custodians for fascinating and coordinating major incidents, incident governance and command, investigations and post incident reports. SOC monitoring is a Business-as-Usual (BAU) and operational function as opposed to cyber onboarding that is usually a project-based time-bounded roles. In some organisations, the SOC analysts can perform varied roles including threat intelligence, intelligence handling, vulnerability management, and threat hunting. Note that these roles ought to be performed by specialist individuals with the appropriate skills regardless of the organisational structure.

---

[1] 'Eyes-on-glass' monitoring is a colloquial to mean people starring on dashboards, computer screens, plasma or projector screens as a means of observing, looking, and detecting an occurrence.

## 2.2    Processes
*SOC processes* in this paper encompass operational guides, local working instructions (LWI), knowledge articles (KA), procedures and operations-level policies. A sample of some SOC essential processes (see Fig. 1) are cyber incident management playbook, incident response process, operational runbook or knowledge articles, joiners, movers and leavers (JML) process, SOC access control policy, security operating procedures (SyOPS) etc.

## 2.3    Technology
The *technology aspect*, as shown in Fig. 1, comprises of the tools that are deployed in a typical SOC, such as SIEM for event analysis, correlation and realtime monitoring; web fraud detection (WFD) to detect web-based transactional fraud, typically for financial orientated SOCs, IDS/IPS to detect and/or prevent intrusions, threat intelligence e.g. malware information sharing platform (MISP - *an open source threat intel feed*) and cyber incident management ticketing system for tracking security incidents tickets, assigning tasks and on-going incidents and issues. There are myriad of SOC tools, but the aforementioned ones in this paper are core and essential.

The SIEM market is very mature with well-established products and a set of criteria to assess their offerings, e.g. Gartner SIEM Magic Quadrant (Gartner, 2018). Mainstream tools range from leaders IBM QRadar and Micro Focus ArcSight to the niche players such as AT&T Cybersecurity, FireEye (Onwubiko, C., and Ouazzane, K., 2019a).

A notable misconception is that many people procure SIEM tools and therefore believe they now have a SOC. This is absolutely incorrect. The tools, when setup properly, will no doubt help the SOC to perform its functions better, provided the 'the challenging' task of onboarding systems, logs, applications and networks to the SIEM is completed, including having the correct parsers, plugins or API (Application Programming Interface) to ingest events from disparate log sources e.g. firewall, routers, applications, intrusion detection systems (IDS) etc. and also, the ability to ingest network-wide information such as flow events and threat intelligence information to detect emerging and inflight incidents (Onwubiko, C., 2015, 2017, 2018).

A SOC must have the appropriate policies and processes to allow them to react swiftly to a cyber incident. For example, a SOC must have a cyber incident management playbook to respond to incidents and coordinate significant cyber incidents (Onwubiko, C., and Ouazzane, K, 2019b), they should have other operating procedures such as SyOPs, cyber recovery process, incident response process and reporting and escalation procedures, at the minimum.

### 2.4 Human-in-the-loop

The people and process aspects, in our opinion, are the fundamental difference between a SOC and a SIEM. We argue that SOCs should have human-in-the-loop; even with artificial intelligence (AI) and machine learning (ML) embedded endpoints and point solutions deployed in the SOC to better and faster detect threats, yet the need for human-in-the-loop cannot be overemphasized. For example, we rely on SOC human operators to make decisions on the cause of action (CoA), not just of technology decision, but holistic decisions that encompasses social, human, financial, risk, reputation and otherwise. We depend on SOC human operators to conduct cyber incident management and to invoke governance and cyber incident and security breach commands, and of course, reliant on technology for pace and precision.

SOC operators (a.k.a. Analysts, Administrators, Incident Responders, Threat Hunters, Forensic Investigators etc.) perform several roles ranging from continuous monitoring, detection, alert triage to threat hunting, incident response and cyber forensic investigation. They continuously monitor business assets and services by leveraging the capabilities offered by tools and technologies deployed in the organisation's estate e.g. SIEM, WFD, Identity and Access Management, IDS, IPS, Anti-malware, firewalls etc.

Automation, orchestration and robotic process automation (RPA), machine intelligence, machine learning and artificial intelligence do play a part, and will continue to play a role in providing power, pace and precision to SOC operations, process and operational efficiencies, but it is best when they are collaborative, cooperative and complementary with humans.

Cybersecurity operations have become increasingly reliant on automation and machine intelligence, and this trend has gained pace in recent years, and will continue at greater pace in the nearest future. This is because automation and machine intelligence-based technologies are becoming readily available and affordable, and are getting better at detecting cyber incidents, especially endpoint and network-based anomalies. For example, machine intelligence-based detection of malware and botnet at enterprise level is now possible (Kidmose, E., 2018).

Unfortunately, process-based automation and human intelligence are complex and hard to codify to machine intelligence, and these are the aspects of cybersecurity operations monitoring that will still be delivered via human analysts, at least for the time being. According to (Guerra P., and Tamburello, P., 2018) "continuous monitoring and detection will remain part of the cybersecurity operations process for the foreseeable future".

## 2.5     Outsource vs. Insource SOCs

The drive to 'outsource' everything was met with 'bring everything back in house' a couple of years ago, and recently, we observe that most companies now operate a hybrid managed SOC model. This is the case, for example, where a framework exists for organisations to outsource some aspects of the SOC service e.g. the continuous monitoring aspects (a.k.a. operations security monitoring) responsibility to a supplier organisation while incident management remains their accountability.

There are many reasons for outsourcing SOC function to supplier organisations, the two main reasons are:
   a) The supplier organisation is tasked to do "the heavy lifting and shifting" – a perception that the expertise to run a functional SOC is readily available in the supplier organisation, hence it is believed that the supplier organisation is by far better to run and maintain a SOC service, while the client organisation becomes responsible for security incident management, escalation and decision making as the overarching risk owner.
   b) Most client organisations work 9am to 5pm, therefore, client organisations prefer to leverage the 24x7[2] SOC service operated by the supplier organisations, a preference many client organisations believe to offer cost saving and efficiency in human resource.

## 2.6     Cloud vs. On-Premise vs. Hybrid SOCs

Cloud-based SOCs have become a new phenomenon. It is the consolidation of an organisations' monitoring capabilities centrally in a Cloud Environment. This encompasses the *hosting* of the SOC's Central Log Collection Infrastructure or Platform, and the tools and technologies they will use to aggregate, collect, collate, curate, process and analyse log, events and intelligence in the Cloud.

---

[2] 24x7 means 24 hours in a day and 7 days in a week.

A *SOC Log Collection Platform* is the central repository where logs, events, network and application metrics are stored, for processing (that is, parsing, normalization, correlation, and alerting) either in realtime or later (e.g. batch processing or non-realtime processing). This then allows analysis and cross correlation of the network information (e.g. flow, packets) and logs and events information in order to detect indicators of compromise (IoC) in packet payloads, logs and events such that alerts can be triaged, and incident response followed in the event of a security breach.

Cloud-based SOCs are becoming popular and increasingly attractive for the following reasons*:*
  a) There are several Cloud-based monitoring tools and technologies that complement SIEM tools in the Cloud. The prevalence[3] of such tools (e.g. CloudTrail, Cloud Watch, Guard Duty, Azure Monitor, Network Watcher, Log Analytics, Application Insights etc.) across the different regions and among the three known and popular Cloud Providers means monitoring of services can be accomplished as quickly as possible, now in *minutes* not *days*, which is the case with on-premise onboarding.
  b) The readiness and availability of APIs and their interoperability means automation and integration, which in traditional SIEM posed several problems are nonetheless irrelevant in Cloud.
  c) Availability of Cloud versions of the SIEM, that is, SIEM providers have built Cloud-based versions of their offering, making their products readily available in the Cloud. E.g. Log analytics, Splunk, ArcSight, ManageEngine, etc all have their cloud offerings in Amazon Web Service Cloud (AWS, 2019).
  d) Pre-Integration and integration with other associated tools are already done by the Cloud provider, and where it has not been done by the provider, it is easy and straightforward. For example, most cloud-based SIEM have pre-integration with tools such as identity and access management, vulnerability management, threat intelligence and anti-malware etc. This means that the SOC can leverage the existing pre-integration to deploy services quicker.
  e) The ability to flex up/down compute and infrastructure in a 'pure' Cloud environment, plus the low cost of entrance (e.g. pay-as-you-use nature of Cloud) has made Cloud-based SOC (a.k.a. Virtual SOCs) a popular and increasingly attractive preposition.
  f) Deliverability and timeliness - Cloud-based SOCs can be built, setup and fully functioning in days as opposed to the more traditional On-Premise SOCs, that take months, at least, to build and setup.

---

[3] Cloud-based monitoring tools are available depending on the Cloud Environment you host your services, e.g. CloudTrail, CloudWatch, Guard Duty are AWS solutions, while Azure Monitor, Network Watcher, Log Analytics and Application Insights are available in MS Azure cloud.

# 3  WHY ARE SOCS IMPORTANT?

The drivers for establishing SOCs are not only driven by *business requirements*, or necessitated by *governance and compliance requirements*, but also, on demonstrable *active risk reduction*.

As an integral part of *business function*, SOCs enable the organisation *fulfil* its business responsibilities and *support* its cyber security strategy. For example, business requirements are underpinned on the appropriate functioning of business services, e.g. *being secure*, *being available to legitimate users of the systems*, and *being integral and trusted*. By using the SOC to continuously and protectively monitor controls (technical, process, policy and procedural) and critical and prioritised business assets of the organisation can the business meets its overarching requirements. Further, it also supports the organisation to meet its business obligations, e.g. business continuity, communication strategy and communications readiness. For example, the SOC's incident management playbook, and majority incident handling protocol should align and inform the organisation's communications readiness in the event of a significant cyber incident or security breach.

As a *compliance requirement,* SOCs are used to fulfill regulatory, governance and legal compliance and directives, for example, regulatory compliance to the payment card industry data security standard (PCI DSS), or compliance to information security compliance, e.g. (ISO 27001), Network and Information Systems Directive (NIS), General Data Protection Regulations (GDPR) etc.

As a mechanism for *active risk reduction*, SOCs are utilized to measure and report key performance indicators (KPIs), such as the percentage of the estate or ecosystem being monitored, systems or critical systems being monitored, performance of continuous vulnerability scans, health hygiene of the IT estate, and progress of any ongoing security incidents and breaches etc. This then provides the organisation with a dynamic active risk picture of their estate.

# 4  CYBER ONBOARDING

Cyber Onboarding follows a set of well-defined processes to onboard a service for cyber security monitoring (see Fig. 2), covering discovery workshop, security monitoring requirements gathering, risk assessment, topology and architecture design, implementation, assurance and security testing, and handover.

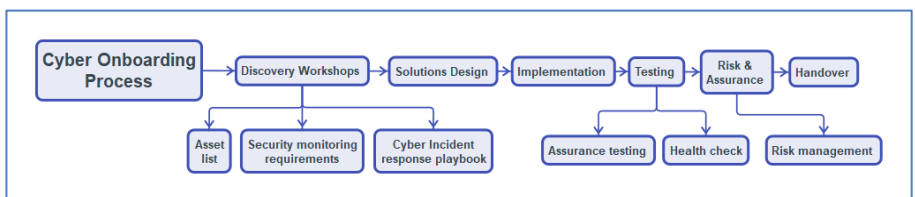

FIG. 2: CYBER ONBOARDING PROCESS

**Note**: Cyber onboarding tasks can be deployed in an agile methodology, which means the entire process lifecycle can be shortened and iterated in small and short sprints; therefore, we are not recommending a lengthy waterfall method. When deployed in Cloud, these processes can be completed in days not months.

These distinct processes are discussed briefly:
   a) Discovery workshops are conducted per organisation, business unit or service to be onboarded to the SOC monitoring platform in order to understand the specific monitoring needs of that organisation, business unit or service such that security monitoring is implemented appropriately to address the unique security monitoring requirements for that department, business unit or service.
   b) Solutions design, architecture and integration patterns are produced based on the organisation's business needs, hosting arrangements, integration requirements, and connectivity options.
   c) Topology map of the existing hosted environments is required in order to allow appropriate monitoring use cases to be developed to ensure that critical assets of the organisation are protected.
   d) The implemented security monitoring solution will need to be tested and assured, and
   e) Finally, the solution is handed over to the SOC to monitor and operate.

Cyber Onboarding is a team in a **SOC function** responsible for ensuring that business services to be monitored by the SOC are appropriately onboarded to the SOC monitoring platform. This means, ensuring that the business services and the underpinning infrastructure and applications within that business area, such as firewalls, servers, desktops and network infrastructures are configured to produce logs and events, and that these events are transported and ingested by the SOC monitoring platform for analysis, correlation, alerting and incident triage.

It is pertinent to note that these assets and applications include containers and microservices, and where logs and events are gathered and monitored, also metrics and state behaviours of the containers and microservices are equally monitored. In some cases, this may require API integration and/or exploitation of container specific mechanisms such as side-car security stacks (SCSS), cybersecurity services, microservices architecture and service mesh (ISTIO) and API management and zero trust monitoring (Chaillan N., 2019).

In some organisations, both the cyber onboarding team and the SOC monitoring and incident management team are the same; however, in this paper, we have presented these teams as distinct but cooperative teams under one management. Hence, the cyber onboarding may not exist as a distinct business unit in most organisations as their duties are performed by the SOC under one accountability business unit. Regardless, the cyber onboarding activities as shown in Fig. 2, must be performed to have a functioning and operational SOC.

These activities include:
- creating design patterns and implementing architecture solutions for any service (existing or new) to be onboarded to the SOC platform for security monitoring;
- ensuring the assets of the business units to be monitored are enabled for logging and events generated by these disparate log sources are ingested and monitored by the SOC;
- enabling the right parsers and plugins so that logs are normalised[4] and forwarded to the SOC platform;
- ensuring that a transport mechanism exists for conveying logs, metrics, events, messages and flows from disparate environments to a central log collection, aggregation and analysis point for the SOC monitoring platform.

These include:
a) Ingest mechanisms: This is a method to ensure that the different and disparate log types generated by the vast array of log sources in the monitored estates are appropriately ingested, normalised and analysed by the SIEM platform. This means ensuring that an ingest mechanism exists e.g., agentless, parser, API and plugin (see log source types in Table 1) for the appropriate log type and format; otherwise, custom parsers must be developed. Custom parsers are especially important for ingesting proprietary logs whose schemas do not comply or conform with appropriate and known standards, e.g. logging standards such as the IETF RFC 5424 format[5].

---

[4] Normalisation is a process of using a consistent schema to process data, events or logs in exactly the same way so that meta-data types are stored on the same columns, for optimised querying and database performance

[5] RFC 5424 – The Syslog Protocol, https://tools.ietf.org/html/rfc5424

22TABLE 1: MONITORING METRIC AND FORMATS (ONWUBIKO, C. 2018)

| S-N | Log Source Type | Log Source Example |
|---|---|---|
| 1 | Events and logs | Raw log, Alert, Event, Windows events, Syslog, Alarm |
| 2 | Network Information | Heartbeat, Flow, Session, Trap |
| 3 | Structured Digital Feed | Scan, Vulnerability Information, PCAP[6], TVM[7], CMDB[8], NVD[9] |
| 4 | Semi and Unstructured Digital | Trace, Manual Input, Wetware |
| 5 | Threat Intelligence | Indicators of Compromise (IoC) |

b) Agent vs agentless: Agent and agentless are both mechanisms to ingest events by the SIEM. Agent-based ingest requires a third-party application or a package of the SIEM to be installed at the end device or endpoint. This is needed, in most cases, when the SIEM tool does not have a matching plugin to ingest logs or events of a particular log source type. For example, windows events do not follow the IETF RFC 5424 standard hence one way to ingest windows events is to install a third-party agent or software at the endpoint to convert windows events to syslog compliant format – this processing of using a third-party software or an agent to ingest logs and event is regarded as agent-based ingestion. The other option is to use agentless method where a third-party agent is not required, instead the SIEM tool accepts native or raw logs or uses API to receives and ingest the events.

c) Design development: The primary function of the technical and solutions architects in the cyber onboarding team is to develop robust and reusable architecture patterns, solutions design and integration patterns artefacts that allow various systems and business services hosted in different locations to be integrated to the SOC monitoring platform, allowing the SOC to securely monitor these services and systems. The created reusable architecture and solutions artefacts are signed-off and approved by the organisation's technical design authorities.

---

[6] PCAP – Packet Capture
[7] TVM – Threat and Vulnerability Management
[8] CMDB – Configuration and Management Database
[9] NVD – National Vulnerability Database

d) Implementation and testing: The design artefacts need to be implemented and tested. Testing can be carried out by other specialist teams, however, this activity should be coordinated through the cyber onboarding team, since they are the project-based arm of the SOC. Testing should not only include assurance testing, but also, security testing such as IT health checks, penetration testing and vulnerability scanning and testing. This is done so that any vulnerability (intrinsic or extrinsic) are mitigated prior to go-live. Since IT health checks are carried to establish intrinsic and extrinsic cyber hygiene of the solution, then it is best to be conducted by an external or independent provider (this is to avoid bias), however, the continuous vulnerability and threat management should still remain an in-house activity.

e) Tagging framework: This is a process of tagging events from specific business services as a way of distinguishing and separating services and this is particularly important in a multi-tenant and multi-customer SOC service, where incident response and escalation maybe different for each business services. Tagging is not only used to differentiate services, but also useful to manage business services with overlapping IP addresses, and where name resolution is not working properly.

f) Alerting and tuning: This is a process of improving the reliability of the service by ensuring that 'noise' and false positives are reduced and minimised. This is done by filtering out known noise on the monitored environment to improve both performance and reliability. The purpose of tuning is to baseline the service so that SOC alerts/alarms are reliable and trustworthy. Tuning do take time and could be considerably longer depending on size, scale and complexity of the SOC platform. On the average, it is common to allow three to six months for this.

g) Network groupings: This is a process of customising networks and subnets into their appropriate business areas, functions and groups to allow for quicker identification of incidents to affected business areas and networks.

h) Content development: This is a process of setting up some of the SOC monitoring artefacts such as rules, filters, use cases, queries and dashboards etc. Monitoring content is important as different business services may face unique risks and concerns; therefore, it is essential that the use cases are adapted to address their respective concerns and risks.

i) Report development: This is a process of creating both generic and custom monitoring reports for each business area and business service being monitored. Reports are used for many purposes, e.g. to

24assess the performance of the SOC service, benchmark the SOC service, review service and operation level agreements (SLA/OLA), key performance indicators (KPI), and most importantly, to measure the return on security investment (RoSI). Cyber metrics such as report against the risks mitigated, report on threats prevented or incidents encountered can be useful barometers to assess RoSI of the SOC (Onwubiko, C. and Onwubiko, A. (2019)).

## 5   Factors Contributing to SOCs Inefficiencies

We argue that some of the contributory factors to SOC inefficiencies include:

a) *Percentage of the monitored estate* – The percentage of the estate monitored is often disproportionate to the coverage of the organisation. We believe that in some organisations, less than 5% of their ecosystem are monitored by their SOCs.
b) *Quality* – Most SOCs provide only a general-purpose basic security monitoring, using 'out-of-the-box SIEM rules' and offering a one-size-fits-all use case. Only very few create custom use cases or bespoke use cases.
c) *Process Maturity* – Most SOCs lack the necessary processes, procedures and local working instructions required to operate the service efficiently. For example, processes for handling cyber incidents, incident management playbooks, etc are found to be non-existent in some SOCs, and where processes exist, they are not regularly updated.
d) *Skilled Resource* – Most SOCs have resource capability and capacity issues (e.g. lack of trained and highly skilled analysts) to run the SOC, this may be related to the global skills shortage in Cyber security, and, retainership of skilled professionals, which is equally a challenge.
e) *Lack of Standardisation* – SOCs have varying perceptions across industry and government. Each organisation has its own understanding of what a SOC should do ((Schinagl, S., et al., 2015), (Onwubiko, C. and Ouazzane, K., 2019a)).

## 6   CSOC Challenges

To build an **effective SOC** takes time, especially one for a large enterprise, such as a government department or financial institution. It is a project that is often dependent on a number of factors, e.g. technical, programmatic, commercial, logistic and organisational. For instance, the footprint of the estate to be monitored, the number of hosting environments to be monitored, size, coverage and complexity of the organisation, the quality of monitoring required and the size of the project workforce, structure and organisation – internal, external, suppliers and partners, procurement frameworks and budget etc. That said, it can be accomplished relatively quickly these days by leveraging cloud computing.

To understand the challenges, we employed a proven methodology – **the reframing matrix**.

The reframing matrix (Mindtools, 2018), created by Michael Morgan (Morgan, M. 1993), is a tool for critical reflection, insight and innovation. An ideal tool for analysing organisational issues from various perspectives that then allows the problem to be viewed from multi-stakeholder perspectives and viewpoints encouraging issues to be seen from different lens, opinions and insights.

| CSOC Perspective | Client Perspective |
|---|---|
| • Not enough feeds to monitor<br>• Poor quality of onboarding e.g. missing parsers & plugins<br>• Incomplete documentation<br>• Too many dashboards | • Lack of funds<br>• Wrong funding model<br>• Lack of return on security investment (ROSI)<br>• Time consuming |
| **Cyber Onboarding is 'Broken'** | |
| • Lack of SMT support<br>• SOC maturity<br>• Onboarding is complex<br>• Dependency issues e.g. lack of business resource | • Too expensive and inconvenient (painful)<br>• Lack of clarity<br>• Governance<br>• Structure |
| Onboarding Perspective | SMT Perspective |

FIG. 3: CYBER ONBOARDING REFRAMING MATRIX

As a problem-solving tool, the reframing matrix uses the four perspectives (4Ps) for insights, viewpoints, interests and concerns. Each quadrant of the matrix is a perspective. The problem to be solved is placed at the centre of the matrix, and opinions, views and concerns are then sought from the respective stakeholders. Based on the different views, solutions to the problems are obtained. It is pertinent that the stakeholders (4Ps) are selected based on their relevance and importance to the problem domain since the strength of the reframing matrix lies on the fact the different stakeholders with different experiences approach problems in different ways.

Our application of the reframing matrix to cyber onboarding is as shown in Fig. 3. First, we put the question been assessed in the middle of a grid. We use boxes around the grid for the different perspectives. Each perspective represents a stakeholder group consulted in the assessment. The 4Ps are the Onboarding Team themselves, the CSOC team, the Client and the Senior Management Team (SMT).

Using the reframing matrix to identify the challenges faced by cyber onboarding (as shown in Fig. 3), we identified **16 different issues** from **four perspective**, namely (clockwise):
- **Onboarding perspective** – as the function responsible for onboarding services for different clients and business units, they deal with the day-to-day fallouts and know the issue best, however, from a unique perspective.
- **CSOC perspective** – as the custodian for security monitoring, and people at the frontline' of the SOC service, so it is important that they are consulted for any reliable solution to the cyber onboarding problem to be identified, besides, they are the direct 'customers' of the Cyber Onboarding Team.
- **Client Perspective** – it is important that we consulted the client for a say, after all, they pay and consume the SOC service. If they are not happy then the business case for standing a SOC capability could easily disintegrate.
- **SMT perspective** – these are the senior management team, comprising the SRO, CTO, Directors and Heads of service. SMT are sponsor, fund and are accountable for the SOC service, therefore has an interest and a viewpoint of the problem.

The 16 issues identified are briefly explained.
From *Onboarding perspective*, they feel that lacked SMT support on a couple of organisational and process issues. They feel SOC is not mature in their operations and skillsets. There is a sense of acceptance that cyber onboarding is indeed complex and complicated, and there are a number of dependencies hindering progress.

From *CSOC perspective*, they feel they are not provided with enough information feeds to monitor. So the onboarding team are not onboarding systems and services quick enough. There is quality issues and incomplete documentations provided to them, which then impacts how quickly they can react, and also, they feel there are many screens to monitor.

From *Client perspective*, there is appreciation of lack of funds. So they do not have funds to pay for the SOC service, and they feel they should not have to pay for a SOC service operated in-house, therefore the funding model is not appropriate. They said there is nothing to show for security monitoring even when it is enabled because they do not receive regular reports or KPIs for the SOC service, and they feel cyber onboarding is very time consuming to rollout.

From *SMT perspective*, they feel cyber onboarding is costing them far too much, hence it is an upscale project. They feel that the metrics and progress they receive from the onboarding team is not clear most times, and that the governance and structure between CSOC and Onboarding teams should be improved.

Following the reframing matrix analysis (see Fig. 3), we conducted a further assessment to see if some of the viewpoints could converge. The 16 viewpoints are now consolidated to 8 key factors that make cyber onboarding challenging and often perceived to be 'broken', as follows:

### A) Complexity

If cyber onboarding is simple then establishing a functioning SOC would not have been so difficult, unfortunately, this is not the case. The process to onboard a service is straightforward in principle (see Fig 2) but often challenging in practice. For example, a service to be onboarded may be hosted in multiple locations and comprising a myriad of different log sources, across the stack, ranging from physical, network, operating systems, middleware, databases to applications. In addition, for a cloud service, this may include hypervisors and/or containers, which also need to be monitored. Each of these stacks will need to be monitored to have a truly complete service onboarding. The problem is that many of these stacks produce logs and messages in varying formats (see Table 1) most of which are non-compliant with the IETF RFC 5424 standard, and a couple may include proprietary formats, especially applications coded in non-compliant formats, therefore the mechanism to ingest and normalise these events is not so trivial. All of these contribute to the complexity, complication and convolutedness.

Additional factors contributing to complexity include A1-A3:

### A1) Architecture designs and patterns

SOC design and architecture is not a one size fits all. Each service onboarding requires a unique design, and at best may leverage existing patterns which will still need to be adapted and implemented, and at worst, a new set of designs are to be produced. The design requirements may be different to the overall design of the SOC monitoring platform itself, therefore, each service to be onboarded will need its own design and solutions architecture, which may utilise existing network connectivity or the provisioning of a new network connectivity to transports logs, events or messages of the onboarded business services to the SOC platform for analysis, correlation and cyber incident triage. The network connectivity (local area networks included) may require a form of wide area network, routing, and security controls enabled to ensure that appropriate policies such as access controls, security groups, blacklisting and firewall policies are correctly implemented.

### A2) Risk assessment

Each business services to be monitored has its own risks or concerns for why it needs security monitoring. For example, a bank implementing security monitoring for their online banking system may do so in order that the SOC will monitor its online bank transactions, hence the risks or concerns are about monitoring of their online banking transactions and ensuring the right customers and correct payments are made; however for a government department responsible for immigration or issuance of national passports, their risks and concerns for security monitoring is obviously different. Here, their concern is to ensure that national passports are only issued to legitimate citizens, that passports are not flaunted on 'black market', and illegitimate documents are not used to obtain national passports. Security risks and concerns are bound to be different based on business functions for different corporations, institutions and government departments. These unique risks and concerns will need to be turned into security monitoring use cases and policies. This process requires niche skillsets, not trivial, and adds a layer of complexity, too.

### A3) Security monitoring requirements

As organisations' business offerings and services are different so are their security monitoring needs. Security monitoring requirements will differ among departments, business units and services, therefore onboarding of each department, business unit or service is bound to be subtly different. While onboarding may follow a fairly straightforward process, however, each business services onboarding requires unique set of solutions ranging from architecture pattern to monitoring use cases.

Take two UK Government Departments for comparison. The Department for Work and Pensions (DWP) for example, their primary responsibility to the UK citizens and government is social welfare to UK citizens in the form of housing allowances, job seekers' allowances etc. to appropriate UK citizens, and on a timely manner. Conversely, HM Revenue and Customs (HMRC) is responsible for collecting taxes e.g. VAT, annual returns, PAYE, customs etc. from citizens and corporations, hence the former's cyber security monitoring need is focusing on ensuring appropriate social welfare arrangements are paid to suitably qualified citizens while the latter ensures and enforces taxes are received from citizens and corporations. Of course, their security monitoring requirements are different and predicated on their business obligations. This goes to demonstrate again that security monitoring and cyber onboarding is not a one size fits all proposition. This uniqueness and tailoring of the cyber onboarding deliverables per business service onboarding adds a layer of complexity and intricacy.

### B) Strategic Support

SOC, like every organisational cyber security programme, has a slim chance of success without strategic support from the senior management teams (SMT). Strategic support is particularly fundamental with SOCs because of its remit, since it serves both as a *horizontal business function*, and as a *compliance mandate*. Without strategic support, SOC will be unable to perform its role of compliance, audit and regulations.

One of the main challenges facing SOCs is having appropriate authority to conduct protective and security monitoring across an entire organisation if SMT have not lend their support and approval. SOC is a horizontal business function, meaning it should be instituted to serve all business units of an entire organisation and should have the prerequisite authority to perform audit, security compliance checks and as an enabler to drive continuous security improvements across the organisation. This is important since cyber-attacks can be exploited from any aspect of the organisation and may use a weakness in one aspect as a channel or conduit to exploit other parts of the business. Hence, SOCs must be empowered, as monitoring custodians, to perform its duties accordingly.

### C) Funding Model

SOC is an upscale project, requiring the procurement and implementation of a myriad of cyber tools, such as SIEM, intrusion detection systems, flow analyser, transaction monitoring (web fraud detection), threat intelligence and possibly user and entity behaviour analytics (UEBA) etc. These tools can be expensive, including software licenses and professional services costs. In addition, the SOC needs facility – the physical operating environment, and human resources to operate and monitor the service and including handling incident response and management. Considering that the project, depending on the organisation's size and scale, may last for a couple of years from start to go-live, and subsequently, the operational people aspect to manage and operate the SOC as normal business as usual (BAU) staff, who must still be costed, then, it is essential that the right funding model for the SOC exists.

The absence of appropriate funding model is likely to impact the success, or the effectiveness of a SOC. SOCs are a medium to address cyber risk and encourage good cyber hygiene, it is therefore pertinent that SOC's funding model is based around *active risk reduction* as other funding models is likely to encourage 'wrong cyber behaviour'. For example, the 'right cyber behaviour' is to encourage active risk reduction as opposed to risk mitigation approach based on 'low hanging fruit'. The reasons for this are that 'easy and quick wins' do not necessarily mean effective prioritisation and efficient risk reduction, because the 'quick wins' may not yield the same risk reduction. We posit that, based on risk proportionality, monitoring an organisation's asset that is either marked for decommissioning or that is not particularly important to the organisation does not yield the same risk reduction as opposed to monitoring the origination's customer database, or their intellectual property.

Similarly, protectively monitoring a standalone guest WiFi just because the guest WiFi project is funded as opposed to offering the same security monitoring on citizens data based on risk reduction encourages wrong cyber behaviour.

Our proposal to addressing the 'cyber behaviour problem', one we strongly recommend, is to ensure that SOC – here we mean SOC and its composite teams such as Cyber Onboarding – is **directly funded**. We distinguish between **direct** vs **central** funding. **Direct funding**, we define as funding allocated directly by the organisation, usually granted or assigned to a business unit and ringfenced for its purpose alone and secured through a business case. On the other hand, **Central funding**, we define as a type of funding arrangement which is obtained by *collectively levying other business units as a contribution for payment of service they have received, or will receive*, and are often referred to as '**cross-charge**'.

SOCs should be **directly funded** to afford it the autonomy to onboard and monitor services that *actively attribute to actual risk reduction*. Prioritisation of services to be monitored by the SOC must not be decided or dictated solely on the basis that an individual business unit has funds or budget, but because the services to be onboarded are those that will *reduce risk exposure* in the ecosystem and to the organisation as a whole.

The premise for onboarding a service just because the project has funds is totally unacceptable. As this may drive wrong cyber behaviour. Fundamentally, if a SOC is centrally funded, it means it has no choice as to which services it monitors, because it will be underpinned on 'first come, first served'. That is, the SOC will serve those who have contributed or paid for their services and this may mean monitoring services of lesser priority/criticality over those that are significantly critical.

### D) Strategy

Every efficient SOC has a clear strategy underpinned by the organisation's Cyber Strategy. Every organisation should have a Cyber Strategy. An organisation cyber strategy is a blueprint for cyber, business transformation, business enablers, governance, risk and compliance.

Organisation Cyber Strategy should adopt cyber principles that encourage, support and enable business and digital transformation agenda, e.g. digital by default, secure by default, active risk management, active defence, proactive and continuous monitoring, cyber resilience and recovery etc. These are the enablers of strong economic wellbeing, creating an environment where businesses thrive by ensuring that digital technology and its frontier are secure. The UK Cyber Strategy (HMG, 2016), a blueprint for national cyber security strategy, aims to create an environment where businesses are confident, capable and resilient in transformational digital world.

For both national and organisational cyber security strategy to be achieved, investments in SOC, Cyber Programme, Governance, Risk and Compliance (GRC), Personnel and Physical security, Cyber Security Training, Awareness and Education need to occur.

### E) Goals and Objectives

With Cyber and SOC strategies come functional goals and objectives. Functional objectives help to achieve business goals, and both in turn enable the strategy to be achieved. To achieve the SOC strategy, high-level **business goals** which are fulfilled by low-level **functional objectives** must exist. A successful SOC function (comprising people, process and technology) is realised on overarching strategy, business goals and functional objectives.

Using the Cyber strategy discussed in Section 4 as an example, a primary goal of the SOC will be to provide realtime security monitoring across the monitored estates. The rationale for this goal is that a goal must directly support its strategy; therefore, to support the SOC strategy of active defence and digital transformation a key enabler is proactive and realtime security monitoring. Further, a key functional objective to achieve the business goal, will be to ensure that the SOC has trained and capable personnel to operate the SOC (i.e. towards SOC maturity).

For SOC to be successful, it must have clear set of goals and objectives that support its strategy, and the wider Cyber Strategy.

### F) Governance and Onboarding Prioritisation

Every organisation should have governance boards, well-defined governance structure, and clear delineation of roles and responsibilities. At a strategic level, there should be a Cyber Governance Board accountable for Cyber. Membership to this board should include the following, at the very least, Cyber SRO, Director of Cybersecurity, Head of GRC, SOC Director/Head, Programme-Level Directors from Business Services. This board should be responsible for deciding on the critical services and systems, through a risk based prioritisation, to be onboarded for security monitoring.

Further, organisational governance structure and hierarchy must be clear so that SOC knows who is in charge with clear point of escalation and reporting. It is important that such structures are communicated not only to the SOC, but also, to the entire organisation. After all, security is everyone's responsibility.

There must be a clear set of rationale based on active risk management for the candidate systems and services to be prioritised. The risk-based prioritisation scheme should take into consideration such metrics as: sensitivity of the assets, criticality of the asset e.g. critical national infrastructure, value of business data it holds e.g. citizens data, business data, national data, cyber value at loss (CVaL), degree of susceptibility of attack, vulnerability of the asset, or that may exist with the controls currently protecting the asset, mean time to restore, disaster recovery targets, cyber response and recovery objectives etc.

### G) SOC Structure and Approach

All the capabilities shown in Fig 1. should sit under one SOC structure. Getting a **SOC structure** right cannot be overstated. It is often the prime causes of an inefficient and immature SOC. The rationale for recommending that all the composite aspects of a SOC sits under one authority is because, it works better and more coherent under one leadership.

If some of the functions, such as Cyber Onboarding were to be under a different structure or authority it will cause friction and fester the perception of 'them' and 'us' mentality, which is needless. Secondly, coherence is key for an effective SOC. That is, the ability to have consistency in processes, administration, methodologies and communication. Communication is important. Information from the SOC to the entire organisation should be concise and consistent.

A SOC structure should support and enable its approach. There are various approaches to operating a SOC, and in this, we are referring to the operating model rather than whether it is outsourced or insourced. The operating model, that is, the SOC operating service hours, for example, 24x7 or 9x5 or 7x7 plus on-call hours. Operating model is governed by business cases determined by the ways of working of all the other stakeholders performing reliant activities either for the SOC or to the business.

Most SOCs operate 24x7 service, which means they work round the clock, 24 hours in a day, 7 days in a week, including Saturdays, Sundays and bank holidays. While some SOCs operate 24x7, this could be arranged as 9x5 plus on-call for after hours and weekend; or 7x7 services complemented with on-call for after hours. Either way, the objective is to have a service coverage that supports the organisation's risk appetite and that are relevant and efficient.

It is pertinent to note that, for example, if a SOC operates 24x7, but some business teams or stakeholder groups are not, then it may make the need for 24x7 SOC ineffective, because if an incident happens during non-working hours and the business teams that are needed to assist with the incident, e.g. networks and infrastructure teams are not 24x7, it then means that the incident will be queued to this team and will be in their queue until when they start work in the following morning. This is not an ideal case and one the puts the effectiveness of the SOC in jeopardy.

SOC operating model must be approved by the SMT based on business case, benefit realisation and business efficiencies. It is important to note that, SOC can operate 24x7 in many formats efficiently as discussed prior.

### H) SOC Maturity

SOC maturity is assessed against many factors, unfortunately, there is no consensus on the factors or criteria that should be used. In this paper, we have carefully selected five generic criteria, we believe should help with operating an effective SOC underpinned on risk reduction, in our assessment. Further, we have also provided a list of some quantitative and qualitative factors that organisations may consider when conducting SOC assessment of their own.

The **generic factors** include:
1. adequate and capably trained staff,

2. robust SOC and Onboarding processes, policies and procedures,
3. appropriately tuned SIEM tool,
4. cyber incident management, reporting and investigation,
5. threat intelligence and threat hunting.

The maturity of a SOC can be assessed on other factors such as **qualitative factors** e.g.
- quality of logging
- how quickly the SOC can recover from a cyber-attack
- how quickly they can respond to a significant cyber incident
- cyber response and recovery readiness
- forensic readiness

On the other hand, SOC maturity can be assessed by **quantitative factors** such as:
- the number of true positives or incidents the SOC detects
- the volume of data analysed in seconds or minutes,
- the number of events processed,
- the number of metrics used in the analysis, e.g. logs, events, flows, PCAP and traps (see Table 1) and
- finally, if monitoring is across the full stack of infrastructure, operating systems, middleware, containers, databases and applications.

Whichever criteria (generic, quantitative, qualitative or a combination of all) are used to assess the maturity of a SOC, there must be rationale for their uses.

### I) Supplier Incentive

As discussed in Section **Error! Reference source not found.**, to build a SOC service often involves multiple stakeholders ranging from internal teams e.g. SOC team, networks and infrastructure teams, to external organisations e.g., suppliers and professional services partners.

For instances, a supplier may be responsible for hosting, another for management of existing legacy services and another for deployment of new services. Whatever their responsibilities are, to deploy a SOC multiple stakeholders are often required. Since the main objective of a SOC is to ensure that all services to be monitored, whether in the supplier environment, hosted applications or cloud-based applications are onboarded, therefore, the SOC will deal with a range of multiple stakeholders and should have a plan to incentivise suppliers and delivery partners in order that the desired outcomes are achieved.

Supplier incentives could be by way of communication to the supplier community of the SOC strategy, and the need for cooperation in order for all assets to be onboarded. This may include *change notices* and *contract change notices* (that is, payment related change notices), impacting and assessment processes that are lean and workable. In addition, supplier incentives may take other forms of collaborative frameworks or memorandum of understanding, such as co-location agreements or deployment of third-party applications into an existing hosting arrangements or procurement of new contractual arrangements.

# 7      RECOMMENDATIONS & CONCLUSIONS

## 7.1      Recommendations

Our recommendations stem from arguments in the preceding sections of this paper. The recommendations are MoSCoW'ed (*Must*, *Should*, *Could* or *Would*) to highlight importance, as follows:

a) An organisation **must** have a cyber strategy upon which SOC strategy and other programme-level strategies hinge, such as network operations centre (NOC) strategy, network and infrastructure strategy, programme management strategy etc. The absence of a cyber strategy will mean that there is no coherent organisation-wide blueprint to work toward, and this is likely to lead to standalone, tower-based models that are fragmented, isolated and divergent.

b) A SOC strategy **should** support and enable the organisation's cyber strategy and offer a mechanism to deliver the cyber strategy.

c) Governance, structure and approach **must** exist, and are fundamental to achieving a fit for purpose and functional SOC. It is imperative to have clear delineation of roles and responsibilities and a distinct line of escalation and reporting, as these will build the enabling environment for an efficient SOC.

d) All SOC composite teams as shown in Fig. 1 **should** be under one authority and governance structure as this will enable the SOC to operate much more efficiently. SOC is complex and adding extra layer of complexity by way of segmenting SOC composite teams under different governance may stifle SOC progress and its autonomy.

e) Whether SOC is funded centrally or directly, having its own ring-fenced funds devolved from individually funded projects allows it to make security decision based on risks rather than funding. Onboarding prioritisation or selection of candidate services to be

    continuously and protectively monitored based on funding drives wrong behaviour as we have seen in Section 4. Hence onboarding prioritisation of candidate system to be monitored **must** be based on active risk reduction.

f) Finally, as SOC is both a horizontal business function and compliance mandate, therefore, it **should** be assessed so that business return on investment and return on cyber security investment are measurable. SOC maturity is one way of achieving this and it is pertinent that the organisation is clear on what metrics or criteria they want to use to measure this growth. As discussed in this paper, we have offered three sets of assessment factors including quantitative, qualitative and generic (see Section 4).

## 7.2	Conclusions

The contributions of this paper are fundamental in many ways. It can be seen as a blueprint for how SOC can be built, delivered and operated; offering a guide to rarely discussed aspect of SOC, which is Cyber Onboarding. The 'aspect' of the SOC that brings new and existing services into the SOC so that it can be appropriately monitored. It is equally the aspect of the SOC that produces the necessary collaterals such as the designs, design patterns, infrastructure and systems configurations and parsers that enable services to be protectively and continuously monitored.

In this paper, we conducted a comprehensive review of SOCs, which enable us to provide a detailed account of the factors contributing their some of their inefficiencies, but most importantly we offered a prioritise set of actions or recommendations other SOCs or organisation can do in order to run an effective and efficient SOC.

The various approaches to building a SOC are discussed, and the choices of an out-sourced SOC vs an In-house (insource) operated SOC are explained, and including the differentiation between SIEM and SOC.

Other key contributions are summarized as follows:

1. SOC is a major organisational investment driven by:
    a. cyber security needs of detection, monitoring, response and recovery from cyber-attacks, especially since modern cyber-attacks are emerging, complex and challenging.
    b. compliance mandate to satisfy regulatory and compliance obligations e.g. PCI DSS, ISO 27001 etc.

2. Building an efficient SOC takes time and effort. Organisations must have a roadmap of SOC delivery aligned with capability and maturity. This is so that it can assess its achievements but more so, to be better planned.
3. SOC is not a one-size-fits-all. Even when a SOC is built for a single organisation, business unit requirements will be different, and risks and concerns are likely to be subtly different and hence SOC and security monitoring use cases must be adapted, tailored and relevant.
4. While SOC processes maybe straightforward, however its success is dependent on cooperation from multiple stakeholders, and in most cases suppliers; therefore, organisations that find themselves in a similar model should have an approach to incentivise suppliers and stakeholders in order that their overarching goals and objectives are accomplished.
5. Finally, SOC must have an operating model, and this must be predicated on business case, relevance and wider stakeholders' ways of working. For example, a SOC can operate 24x7 in multiple ways; and of course, should not operate 24x7 if the organisation's business case and risk appetite dictate differently.

# 8   FUTURE RESEARCH DIRECTIONS

We aim to investigate a quantitative research with possibly London Metropolitan University as a stakeholder to prototype some aspects of the findings in this paper. This is so that a demonstrable Minimum Viable Product (MVP) may be realised.

## BIOGRAPHICAL NOTES

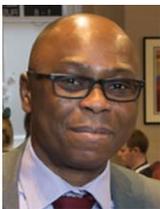

**Dr Cyril Onwubiko** is a Senior Member of the IEEE, and currently the *Secretary – IEEE United Kingdom and Ireland*, Chair of the *IEEE UK & Ireland Blockchain Group;* and Director, Artificial Intelligence, Blockchain and Cyber Security at Research Series Limited, London, UK, where he directs Artificial Intelligence, Blockchain and Cyber Security. Prior to Research Series, Cyril had worked in the Financial Services, Telecommunication, Health sector & Government and Public services Sectors. He is a leading scholar in Cyber Situational Awareness (Cyber SA), Cyber Security, Security Information and Event Management (SIEM) & Data Fusion, he has authored a couple of books including "Security Framework for Attack Detection in Computer Networks", and edited several books including "*Situational Awareness in Computer Network Defense: Principles, Methods & Applications*". Cyril is the Editor-in-Chief of the International Journal on Cyber Situational Awareness (IJCSA), and founder of the Centre for Multidisciplinary Research, Innovation and Collaboration (C-MRiC). He holds a PhD in Computer Network Security from Kingston University, London, UK; MSc in Internet Engineering, from University of East London, London, UK, and BSc, first class honours, in Computer Science & Mathematics. He's a guest lecturer to several UK and European Universities and have successfully examined / assessed several PhD students in the UK and Europe.

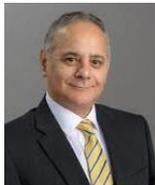

**Professor Karim Ouazzane** is a professor of Computing and Knowledge Exchange, Director of Research and Enterprise, Chair of the European Cyber Security Council (Brussels), and founder of the *Cyber Security Research Centre (CSRC),* London Metropolitan University, London, UK. His research interests include artificial intelligence (AI) applications, bimodal speech recognition for wireless devices, cyber security and big data, computer vision, hard and soft computing methods, flow control and metering, optical instrumentation and lasers. He has carried out research in collaboration with industry through a number of research schemes such as The Engineering and Physical Sciences Research Council (EPSRC), KTP, EU Tempus, LDA (London Development Agency), KC (Knowledge Connect) and more. He has also published over 100 papers, three chapters in books, is the author of three patents and has successfully supervised 13 PhDs. He is a member of the Oracle Corporation Advisory Panel.

## REFERENCE

**Reference to this paper should be made as follows**: Onwubiko, C. and Ouazzane, K. (2019). Challenges towards Building an effective Cyber Security Operations Centre. *International Journal on Cyber Situational Awareness*, Vol. 4, No. 1, pp11-39.